\documentclass[a4paper, 10pt, conference]{ieeeconf}      

\IEEEoverridecommandlockouts                              
\usepackage[pdftex]{graphics} 
\usepackage{epsfig} 
\usepackage{mathptmx} 
\usepackage{amsmath} 
\usepackage{amssymb}  
\usepackage[a4paper]{geometry}

\title{\LARGE \bf
Fuzzy Logic based Autonomous Parking Systems - Part IV: A Multiple-Model Adaptive Neural-Fuzzy Controller}

\author{Yu Wang and Xiaoxi Zhu
\thanks{Yu Wang is with department of Electrical Engineering, Yale University, 06510 {\tt\small yu.wang@yale.edu}. Xiaoxi Zhu is with Singapore Telecommunications Limited {\tt\small zhu.xiaoxi@hotmail.com}}}

\begin{document}

\maketitle
\thispagestyle{empty}
\pagestyle{empty}

\begin{abstract}
In this paper, a Multiple Models Adaptive Fuzzy Logic Controller (MM-AFLC) with Neural Network Identification is designed to control the unmanned vehicle in Intelligent Autonomous Parking System. The objective is to achieve robust control while maintaining a low implementation cost. The proposed controller design incorporates the following control theorems -- non-linear system identification using neural network, fuzzy logic control, adaptive control as well as multiple models adaptation. Such integration ensures superior performance compared to previous work. The generalized controller can be applied to different systems without prior knowledge of the actual plant model. In the intelligent autonomous parking system, the proposed controller can be used for both vehicle speed control and steering wheel turning. With a multiple model adaptive fuzzy logic controller, robustness can be also assured under various operating environments regardless of unpredictable disturbances. Last but not least, comparative experiments have also demonstrated that systems equipped with the new controller are able to achieve faster and smoother convergence. 
\end{abstract}

$\it Key words$--Fuzzy Logic Control, Adaptive Control, Neural Network, Multiple Models Identification, Unmanned Vehicle, Autonomous Parking, Intelligent Transport System


\section{Background}

Unmanned vehicle or autonomous driving has been a hot topic in both academics and industries in the past decade. However, there is a gap between research and application. Despite the fact that many advanced control theorems have been proposed for cutting-edge intelligent vehicle, the conventional approach of PID control remains as the most widely adopted method for industrial applications.(\cite{c1},\cite{c2}) The underlying reason is that PID controller delivers a good balance among control performance, implementation cost and operational complexities. Generally, PID controller gives best performance on linear system with lower order. However, most industrial plants are non-linear in nature and the linearisation leads to high order. Besides, unpredictable disturbances cannot be taken care of in the controller design and may negatively affect the control performance. Last but not least, stability of PID controllers cannot be promised with unknown system parameter and external disturbances.

Recently, Fuzzy Control has gained much attraction as an alternative solution to PID control in industrial applications. Fuzzy Control is a rule-based decision making which simulates the way of human reasoning. Unlike traditional controllers where numerical parameters are used, fuzzy controllers define linguistic variables using fuzzy sets. This underlying fuzzy logic improves robustness in dealing with system uncertainties and external disturbances. The work in \cite{c3} and \cite{c4} demonstrates successful application of fuzzy logic in motor speed control.

Another method to compensate for indeterminacies is to build observers and/or estimators. In autonomous driving systems, road surface friction is a major disturbance which may seriously affect the system performance. In \cite{c5} and \cite{c6}, estimation of friction is obtained through an observer and subsequently the information can be used to complement the controller. One drawback is that sliding mode control does not give optimal performance if the system uncertainties are unstructured. 

Alternatively, indirect adaptive control can be used. As the name indicates, the controller relies on indirect measurement — performance error between actual plant output and desired value —- and adjusts control signal continuously to achieve better outcome. Therefore, it does not require prior knowledge of plant model and can be implemented on different systems operating in different environment. Apparently the computation cost is high for such on-line processing. 

This paper proposes a new scheme for robust control of unmanned vehicle. The control process consists of two parts - plant identification with Neural Network theory followed by a multiple models adaptive fuzzy logic control. It optimize the control performance against uncertainties (unknown plant model, external disturbances, etc) while managing the the implementation complexities.

\section{System Design}

As discussed in previous work \cite{XYZ2012}, \cite{XYZ2013} and \cite{XYZ2014}, there are two major control objectives in an unmanned autonomous driving system -- speed of driving wheel and turning angle of driven wheel, which corresponds to rotational speed of motor and turning angle of steering wheel. Control performance cannot be guaranteed given the unknowns - unidentified plant model and unpredictable environmental disturbances. On one hand, both plants are complicated mechanical devices where an accurate mathematical model is difficult to derive. On the other hand, even if a mathematical model can be used to represent the system (such as an induction motor which can be described by a second order differential equation), it is still difficult to determine the actual parameters which may vary across different instances and operating environment. Last but not least, it requires extensive efforts to derive a robust controller for complex non-linear plants with performance guarantee, such as stability, fast convergence, etc. Not to mention that actual implementation of such a controller is yet another challenge when it comes to real industrial applications.

A two-step approach is proposed in this paper to overcome above-mentioned difficulties so that optimal control performance can be achieved with minimal cost. First of all, neural network is used to identify the plant model through an off-line training process. Secondly, multiple models adaptive fuzzy logic control is used to achieve faster and robust control performance. Comparing to previous work with single or hybrid fuzzy logic controllers, significant improvements results from two design enhancements - additional off-line model identification and optimized on-line adaptation process.

\subsection{Design Approach}

A fuzzy logic controller can be implemented without prior knowledge of plant. As a result of its inherent characteristics, fuzzy logic controllers operates well under different environment and deliver consistent performance regardless of disturbances. Therefore, it can be widely applied at minimal implementation cost.  However, inferior performance such as overshoot, oscillation and slow convergence are inevitable given big variations in plant model (internal) and operating environment (external). 

On-line adaptation of fuzzy logic controller is proposed to improve the control performance -- reducing oscillation in particular. Controller parameters are adjusted on-line by applying Lyapunov Adaptation Law, using performance error (difference between actual output and desired value) as the tuning signal. Further enhancement can be achieved by introducing multiple models adaptation. The controller with m adjustable parameters can be represented as a single point in M-dimensional space. Assuming there is a point representing the optimal controller, the adaptation process can be visualized as the trajectory from starting point to destination. Instead of single starting point, (m+1) origins are selected to define a region which encloses the optimal point. The adaptation can be viewed as a zoom into the optimal point. This approach can be proved to deliver faster convergence so that oscillation can be minimized as well. 

It is obvious that selection of the starting points are critical, otherwise the adaptation process will be prolonged with large overshoot and oscillation. An off-line neural network model identification is proposed to address this issue. As neural network is a generalized approach for plant identification, it can be applied to identify non-linear plant without knowing the actual model. It can also be proved that error between actual plant and identified model is bounded by $\epsilon$. Given the identified plant model, a controller can be designed to generate the control signal so that a desired plant output is achieved given any reference input. However, stability of the derived controller cannot be guaranteed in this case.  Nevertheless, the identified model is only an approximation and the actual operating plant model changes constantly due to unpredictable external disturbances. These two issues can be resolved through multiple models on-line adaptation process discussed in previous paragraph. The approximation will be critical to the initial design of fuzzy controller to ensure faster convergence. 

A detailed design of neural network identification and multiple models adaptive fuzzy logic controller is presented in this section. 

\subsection{Identification of Non-linear Plant with Neural Network}
In previous work  \cite{XYZ2012}, \cite{XYZ2013} and \cite{XYZ2014}, the focus is on designing an robust Fuzzy Logic Controllers to achieve best outcome in autonomous parking regardless of internal and external variations. Those controllers were designed without any prior knowledge of the plant model (unmanned vehicle in this case) as an exact mathematical representation is difficult to derive due to the non-linearity and complexity of the plant. Therefore, the accuracy, stability and robustness of controllers are compromised. In order to improve the control performance, a method using continuous-time neural network is proposed for the purpose of model identification and further ensures the stability of fuzzy-controller described in next section.

Assuming that the plant -- an unmanned vehicle -- can be described by a second-order system with two states $x_1$ and $x_2$, If the controlled plant is driven wheel where a target steering angle is set,  $x_1$ represents the angular position while $x_2$ represents the angular speed. if the controlled plant is driving wheel to achieve a target speed,  $x_1$ represents the current position while $x_2$ represents the speed. Therefore, the plant to be identified and controlled can be described by the following functions:

\begin{equation}
\dot{x}_1(t)=x_2(t),\;\;\;\;
\dot{x}_2(t)=f_0^T(x_1(t),x_2(t))\theta+u(t)
\end{equation} 

where both $x_1$ and $x_2$ are scalar values. $f_0(x_1(t),x_2(t))$ is an unknown non-linear function defined by states $x_1$ and $x_2$, which are both measurable and accessible. $\theta$ is a group of linear parameters that needs to be identified. It shall be noted that $\theta$ is completely decoupled from x. The control signal u(t) is defined as a scalar which is not correlated with function $f_0(x_1(t),x_2(t))$. 

Therefore, the system representation can be further simplified by substituting the vector $[x_1(t),x_2(t)]^T$ with $x$ as below:

\begin{equation}
\dot{x}(t)=f(x(t))\theta+bu(t)
\label{plant}
\end{equation} 
where $x(t) \in R^2$ and $b=[0;1]^T$.

\underline{Parallel identification model:}

\begin{figure}[thpb]
     \centering
     \includegraphics[width=8cm,height=4cm]{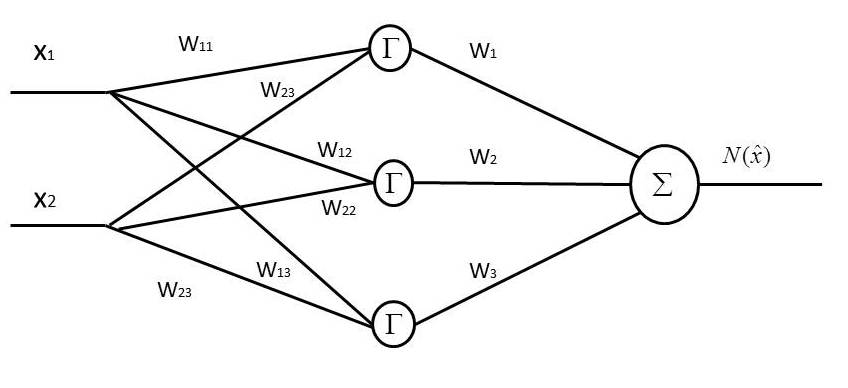}
     \caption{Neural Network Identification Structure}
     \label{neural_structure}
   \end{figure}

Due to the non-linearity of system as described in Equation (\ref{plant}), a three-layer neural network (with one hidden layer) $N(x)$ is proposed to identify the plant model $f(x)$. The neural network structure is shown in Figure (\ref{neural_structure}). Define $\hat{x}(t)$ as the estimated states. Thus the identification system using parallel identification models can be expressed as: 

\begin{equation}
\dot{\hat{x}}(t)=N(\hat{x}(t))\theta+bu(t)
\label{identification}
\end{equation}

and 
\begin{equation}
N(\hat{x}(t))=\sum\limits_{i=1}^3 w_i \Gamma(\sum\limits_{j=1}^2 w_{ij}x_{j})
\label{NN}
\end{equation} 

where $w_{ij}$ represents weight between input layer and hidden layer and $w_{i}$ represents weight between hidden layer and output layer.  The non-linear activation function $\Gamma$ which is applied on hidden layer nodes is chosen as a hyperbolic tangent function 
\begin{equation}
\Gamma=\dfrac{e^x-e^{-x}}{e^x+e^{-x}}
\end{equation}  

In the late 1980s and early 1990s, the theory has been proved that a multi-layer neural-network, even with as few as one hidden layer, can be used to approximate any continuous function on a compact set. The same conclusion and its proof have been provided by numerous researchers independently, including Funahashi\cite{Funahashi1989}, Cybenko\cite{Cybenko1989}, Gallant\cite{Gallant1988} and Hornik\cite{White1989}. Therefore, it can be concluded that neural network is a universal approximation for non-linear functions as long as the activation function is selected properly.

In order to derive the weights in the identification system, an off-line training process is implemented through back propagation method by using identification error as the tuning signal.

The method used here is back propagation with gradient algorithm. Define $f_d=[f_{d1},f_{d2},...,f_{dm}]$ as a group of real f(x) values at different time instances (i.e $f_{d1}=f(x(t_1))$, $f_{d2}=f(x(t_2))$, etc.). The performance criterion to be minimized can be defined as $L=\dfrac{1}{2}\parallel N(\hat{x})-f_d\parallel^2$. 

The adaptation law for adjusting the weights shall be:

\begin{equation}
w_i(k+1)=w_i(k)-\alpha\dfrac{\partial L}{\partial w_i(k)}
\end{equation}
where $\alpha$ is the training speed of Neural Network.

\subsection{Multiple Models Adaptive Fuzzy Logic Controller}

\begin{figure}[thpb]
     \centering
     \includegraphics[width=9cm,height=8cm]{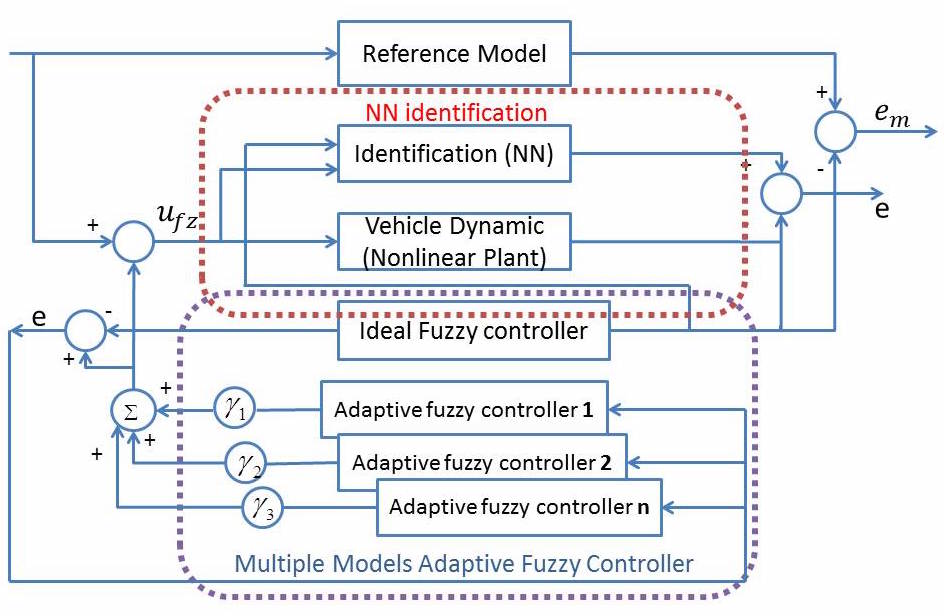}
     \caption{Control Diagram for MM-AFLC}
     \label{controlnew}
   \end{figure}

A fuzzy logic controller can achieve good control performance in spite of variation in plant parameters and external disturbances. By introducing on-line adaptation, it saves a lot of efforts in parameter tuning so that the same controller can be easily used on different plants. The integration of multiple models adaptive control further improves the control accuracy and results in a faster and smoother convergence. Figure (\ref{controlnew}) is the control diagram which illustrates the working principle of the proposed Multiple Models Adaptive Fuzzy Logic Controller (MM-AFLC).

The principal ideas involved in Multiple Models Adaptive Fuzzy Logic Controller (MM-AFLC) are discussed in great details below. The section covers the mathematical preliminaries of adaptive fuzzy control and its extension to multiple models based adaptive fuzzy controller.


\subsubsection{Adaptive Control with Model Identification}
In conventional adaptive control, the most important step is to define the error equation and Lyapunov Candidate Function to design the adaptive controller.

Firstly, define a linear reference model which the plant should follow, the objective is to control the state $x_2$, despite any internal and/or external uncertainties. Mathematically the control objective can be expressed as the model equation below:

\begin{equation}
\dot{x}_m = A_m x_m + b_m r
\label{reference model}
\end{equation}

where r is the reference input, $A_m$ and $b_m$ are reference model parameters, and $\dot{x}_m$ is the desired system states.

In order to ensure the stability of the indirect adaptive control system, neural network model identification discussed in previous subsection is incorporated in the system design. Compared to a parallel identification model shown as in (\ref{identification}),  the improved approach -- series-parallel identification model -- is able to identify and control the plant in a more stable manner.
The series-parallel identification system can be constructed as:

\begin{equation}
\dot{\hat{x}}=A_m\hat{x}+(N(\hat{x})(t)\hat{\theta}-A_m x(t))+b\hat{u}
\label{identification2}
\end{equation} 
 
where $\dot{\hat{x}}$ is the estimated value of the vehicle states x, $N(\hat{x}(t))$ is the neural network estimation of the non-linear function $f(x)$.

The corresponding control signal can be rewritten as

\begin{equation}
{u}^* = k^*(t)x(t)+r
\end{equation}
 
where
\begin{equation}
k^* = {b}^{-1}(A_m-f(x)\theta/x)
\end{equation}

In the process of adaptation, the control signal $\widehat{u}$ will also converge to the desired control signal. The stability analysis is discussed in details below.

Define the control signal as:

\begin{equation}
\widehat{u}(t) = \widehat{k}(t)x(t)+r
\end{equation}

where

\begin{equation}
\widehat{k} = {b}^{-1}(A_m-N(\hat{x})\widehat{\theta}/\hat{x})
\label{linearility}
\end{equation}

By subtracting (\ref{identification2}) from (\ref{plant}), we may have that the error model equation:

\begin{equation}
\dot{e} = A_me + bN(\hat{x}(t))\tilde{\theta}
\label{error_model}
\end{equation}

where e is the control error and $\tilde{\theta}$ is equal to $\hat{\theta}-\theta$. The original non-linear part f(x) in plant is substituted by $N(\hat{x})$ if the off-line identification of $f(x)$ is good enough, such that $N(\hat{x}(t))$ is closed enough to $f(x)$.

Lyapunov method is used here to design the adaptive law that ensures system stability and the Lyapunov candidate function is chosen as:

\begin{equation}
V(e,\widetilde{\theta}) = \frac{1}{2}[e^2+\widetilde{\theta}^2]
\end{equation}

where

\begin{equation}
\widetilde{\theta} = \widehat{\theta}(t)-\theta
\end{equation}

Taking the derivative of V w.r.t t gives:

\begin{equation}
\dot{V} = e\dot{e}+\widetilde{\theta}\dot{\widetilde{\theta}}
\label{lyapunov2}
\end{equation}

Substitute (\ref{error_model}) into (\ref{lyapunov2}) and the derivative can be re-written as:

\begin{equation}
\dot{V} = A_m {e}^2 + \widetilde{\theta} (e^TbN(\hat{x})+\dot{\widetilde{\theta}})
\end{equation}

The system is stable if the Lyapunov function is locally positive definite and its derivative is locally semi-definite negative. To fulfil the requirement the adaptive law is defined such that:

\begin{equation}
\dot{\widetilde{\theta}} = -e^TbN(\hat{x})
\end{equation}

Also from the control equation (\ref{linearility}), the relationship between feedback control signal $\dot{\widehat{k}}$ and the adaptive parameter $\dot{\widehat{\theta}}$ is linear. i.e.

\begin{equation}
\dot{\widehat{k}}=\dot{\widehat{\theta}}=\dot{\widetilde{\theta}} = -e^TbN(\hat{x})
\end{equation}

Since $A_m$ is always negative definite for a stable reference model, $\dot{V} = A_m e^2$ is also non-positive definite.

In addition, the error convergence can be shown by:
\begin{equation}
-\int^{\infty}_{t_0} \dot{V}(e(\tau),\widetilde{A}_m(\tau))d\tau = V(t_0)-V(\infty)=0
\end{equation}

hence:

\begin{equation}
0\leqslant \int_0^\infty (e)^2(\tau)d\tau < \infty
\end{equation}

where $e\in \mathcal{L}^2$

As $\dot{e}$ is bounded, it follows by Barbalat's lemma that $\lim_{t\rightarrow\infty} e(t)=0$

\subsubsection{Adaptive Fuzzy Logic Controller}

The adaptation process with model identification discussed above does not specify the type of controller selected. The biggest advantage is that it resolves the issue with unknown internal parameters -- plant model. In order to achieve optimal control against external disturbance, an adaptive fuzzy logic controller can be proposed. As the plant model of electrical motor can be modelled as a linear system, a small fuzzy rule base is enough for the decision making, where the output fuzzy set is defined to be adaptive, meaning that parameters $\alpha_1$, $\alpha_2$, $\alpha_3$,...,$\alpha_n$ are adjustable.

The defuzzified output of traditional fuzzy controller can be expressed as:

\begin{equation}
u_{fz}(\mathbf{\alpha})=\mathbf{\alpha}^T\mathbf{\xi}
\end{equation}

where $\mathbf{\alpha}=[\alpha_1, \alpha_2,...,\alpha_m]^T$ is a parameter vector and $\mathbf{\xi} = [\xi_1, \xi_2,...,\xi_m]$ is a regressive vector with $\xi_i$ defined as
 
\begin{equation}
\xi_i = \frac{w_i}{\sum_{i=1}^{m} w_i}
\end{equation}

According to Universal Approximation Theorem, there exists an optimal fuzzy control signal such that

\begin{equation}
u^*(t) = u_{fz}^*(\mathbf{\alpha}^*)+\epsilon = \mathbf{\alpha}^{*T}\mathbf{\xi}+\epsilon
\end{equation}

where $\epsilon$ is the approximation error bounded by $\vert\epsilon\vert <E$. The actual fuzzy logic controller with adjustable parameters can be viewed as an identification model of the optimal fuzzy controller, and equivalently an estimator of $u^*$, which can be expressed as

\begin{equation}
\widehat{u}_{fz}(\widehat{\mathbf{\alpha}})=\widehat{\mathbf{\alpha}}^T\mathbf{\xi}
\end{equation}

\subsubsection{Multiple Models Adaptive Fuzzy Logic Controller}

In order to overcome the weakness of low accuracy and slow convergence by using fuzzy logic, a multiple models approach are proposed to further enhance the control performance. It accelerate the adaptation process by restricting the unknown parameters in a convex hull. The following section discusses the improvement plan in extensive details.   

Instead of using single identification model of fuzzy controller, N models are used to identify the desired fuzzy logic controller to improve the accuracy and speed simultaneously.

The N identification controllers $\Sigma_{1}, \Sigma_{2} ,..., \Sigma_{N}$ can be expressed as:

 \begin{equation}
\Sigma_i:       \widehat{u}^{(i)}_{fz}(\widehat{\alpha}^{(i)})=(\widehat{\mathbf{\alpha}}^{(i)})^T\mathbf{\xi}, i \in {1,2,..N}
\end{equation}

where $(\widehat{\mathbf{\alpha}}^{(i)})^T=[\alpha^{(i)}_{1}, \alpha^{(i)}_{2},...,\alpha^{(i)}_{m}]^T$ is the parameter for $i^{th}$ identification controller, their initial values can be chosen to set up a convex hull such that the single identification controller described by (25) can be expressed as a linear combination of N multiple identification controllers, i.e. 

\begin{equation}
\widehat{u}_{fz}(\widehat{\mathbf{\alpha}})=\widehat{\mathbf{\alpha}}^T\mathbf{\xi}=\sum_{i=1}^{m}(\gamma_i(\widehat{\mathbf{\alpha}}^{(i)})^T)\mathbf{\xi}
\end{equation}
for $\sum_{i=1}^{m}\gamma_i$=1 and $\gamma_i \geq 0$. This indicates that the desired fuzzy controller will fall in the convex hull of the N identification controllers.

Also the output error for each of the $i^{th}$ identification controller is expressed as:

\begin{equation}
\widetilde{u}^{(i)}_{fz}=((\widehat{\mathbf{\alpha}}^{(i)})^T-\mathbf{\alpha}^{*T})\mathbf{\xi} = (\widetilde{\mathbf{\alpha}}^{(i)})^T\mathbf{\xi}
\end{equation}

The system model and its estimator can be re-written as:

\begin{equation}
\dot{x} = f(x)\theta + b u^*
\end{equation}
\begin{equation}
\dot{\widehat{x}}^{(i)}=A_m\widehat{x}^{(i)}+(N(\hat{x})(t)\hat{\theta}-A_m x(t))+b\widehat{u}^{(i)}
\label{fuzzy identification model}
\end{equation}

where $\widehat{u}^{(i)} = \widehat{u}_{fz}^{(i)}+\eta^{(i)}(t)$. $\eta^{(i)}(t)$ is the transient error between $\widehat{u}^{(i)}$ and $\widehat{u}_{fz}^{(i)}$ and it can be proven that $\eta^{(i)}(t)$ is bounded. $\widehat{x}^{(i)}$ is the resulted system states given the controller is $\widehat{u}^{(i)}$

Subtract (\ref{fuzzy identification model}) from (\ref{reference model}), the following is obtained:

\begin{equation}
\dot{e}_m^{(i)} = A_m e_m^{(i)} + b(\widehat{u}_{fz}^{(i)}+ \eta^{(i)}(t) -\epsilon^{(i)})
\end{equation}

where $e_m$ is defined as $x-x_m$
Choose a Lyapunov candidate function as:
\begin{equation}
V = \frac{1}{2}({e_m^{(i)}})^2 + (\widetilde{\mathbf{\alpha}}^{(i)})^T\widetilde{\mathbf{\alpha}}^{(i)} \geqslant 0
\end{equation}

therefore, the derivative of Lyapunov candidate is:

\begin{equation}
\dot{V} = e_m\dot{e}^{(i)}_m + (\widetilde{\mathbf{\alpha}}^{(i)})^T\dot{\widetilde{\mathbf{\alpha}}}^{(i)}
\end{equation}
\begin{equation}
\dot{V}=A_m (e_m^{(i)})^2+(\widetilde{\mathbf{\alpha}}^{(i)})^T(e_m^{(i)} b \xi +\dot{\widetilde{\mathbf{\alpha}}}^{(i)})+e_m^{(i)} b(\eta^{(i)}(t)-\epsilon^{(i)})
\end{equation}

To ensure stability, the adaptive law is chosen such that the second term on the right side of (34) is always zero. i.e.

\begin{equation}
\dot{\widetilde{\mathbf{\alpha}}}^{(i)} = -e_m^{(i)} b \mathbf{\xi}
\end{equation}

In addition, define
\begin{equation}
\eta^{(i)}(t) = -Esgn(e_m)
\end{equation}

which ensures that $e_m^{(i)}b\eta^{(i)}(t)-\epsilon^{(i)}$ is always negative. Therefore, $\dot{V}$ is always negative and the system is stable.

Though the adaptive law discussed above ensures the stability of the controller, the parameter information of each identification controller is still used separately during the control process. One of the idea to further enhance the controller's performance is to combine the information of multiple controllers instead of just using one.  Intriguing by this, the controller will be further implemented based on the convex combination property as shown in (27)
By subtracting (25) from (27), it gives us:

\begin{equation}
\sum_{i=1}^{m}\gamma_i((\widehat{\mathbf{\alpha}}^{(i)})^T-\widehat{\mathbf{\alpha}}^T)=0
\end{equation}

where $\sum_{i=1}^{m}\gamma_i=1$. For further simplicity, define $\widehat{\mathbf{\alpha}}^{(i)}-\widehat{\mathbf{\alpha}}=\phi_{i}$, $\Phi=[\phi_{1}-\phi_{m},\phi_{2}-\phi_{m},...,\phi_{m-1}-\phi_{m}]$ and $\gamma=[\gamma_1,\gamma_2,...,\gamma_{m-1}]$. Thus, (46) becomes

\begin{equation}
\Phi^T\gamma=-\phi_{m}
\end{equation}

multiplying both sides by  $\Phi$, we have

\begin{equation}
\Phi\Phi^T\gamma=-\Phi\phi_{m}
\end{equation}

Therefore, the multiple models adaptive fuzzy logic controller can be constructed by using the differential equation to estimate $\gamma$, which is also the weight of each identification controller in the linear convex combination. The control law is:

\begin{equation}
\dot{\widehat{\gamma}}(t)=-\Phi\Phi^T\widehat{\gamma}(t)-\Phi(t)\phi_{m}(t)
\end{equation}

Hence,

\begin{equation}
\dot{\widetilde{\gamma}}(t)=-\Phi\Phi^T\widetilde{\gamma}(t)
\end{equation}

where $\widetilde{\gamma}=\widehat{\gamma}-\gamma$ is the estimation error.

To examine the stability of the proposed MM-AFLC, using a Lyapunov Candidate function given as:

\begin{equation}
V(\widetilde{\gamma})=\frac{1}{2}\widetilde{\gamma}^T\widetilde{\gamma}
\end{equation}

Which follows that

\begin{equation}
\dot{V}(\widetilde{\gamma})=-\widetilde{\gamma}(t)^T\Phi\Phi^T\widetilde{\gamma}(t)=-||\Phi^T\widetilde{\gamma}(t)||^2 \leq 0
\end{equation}

Hence, since the derivative is non-negative, (41) is stable and $\widetilde{\gamma}(t)$ is bounded.

\section{Simulation}
Comparative experiments are conducted with Matlab simulation to conclude the performance improvement, comparing with three other fuzzy controller based approaches-FBOS, HFC and HFC-FTD, which are  proposed separately in \cite{XYZ2012}, \cite{XYZ2013} and \cite{XYZ2014}. These experiments are conducted in a parallel manner such that all external environment parameters are controlled to be the same. The ground friction is chosen to be much larger than normal value in order to test the robustness and performance of each controller. Different controllers are applied on the same plant model to ensure the fairness of the comparison.

In order to test the performance of FBOS, HFC and HFC-FTD, the controller parameters need to be configured before the test. The parameter selection for FBOS and HFC is based on trial experiments(\cite{XYZ2012}, \cite{XYZ2013}), while that for HFC-FTD is through a data training process as presented in \cite{XYZ2014}. For our new proposed method, an off-line training process for the non-linear plant using Neural Network need to be performed beforehand as illustrated in Section II.

Once the controllers parameters are configured properly, the experiment may start.  The entire parking process are simulated and recorded for each individual controller. In addition, the simulation is recorded such that the gap between every two frames is kept as a constant--0.1 second. It ensures that the performance and speed of different controllers can be easily compared. In other words, a more scattered path corresponds to a faster parking process.

\begin{figure}[thpb]     
     $
\begin{array}{cccc}
    \includegraphics[width=4cm,height=3.5cm]{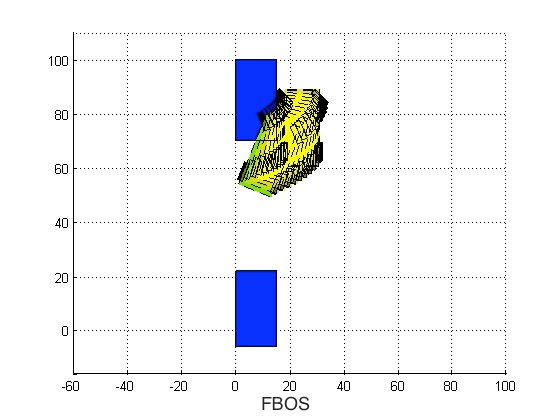}&
    \includegraphics[width=4cm,height=3.5cm]{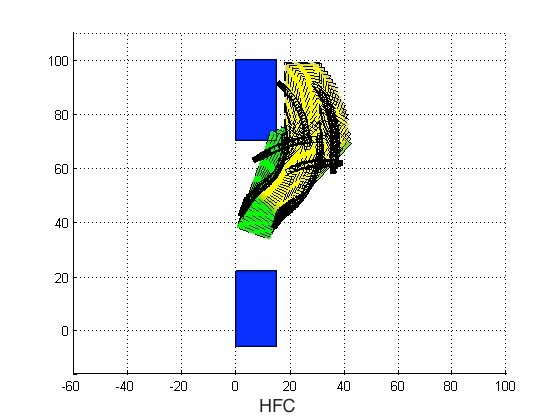}&\\
    \includegraphics[width=4cm,height=3.5cm]{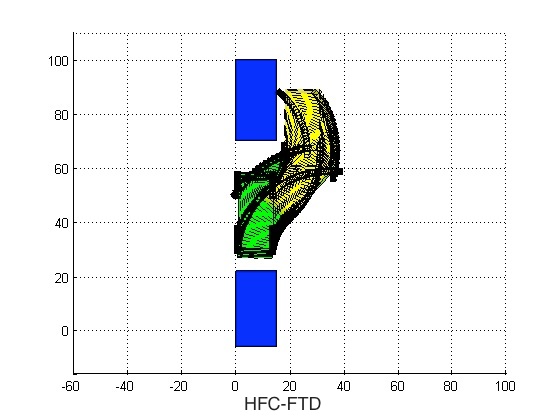}&
    \includegraphics[width=4cm,height=3.5cm]{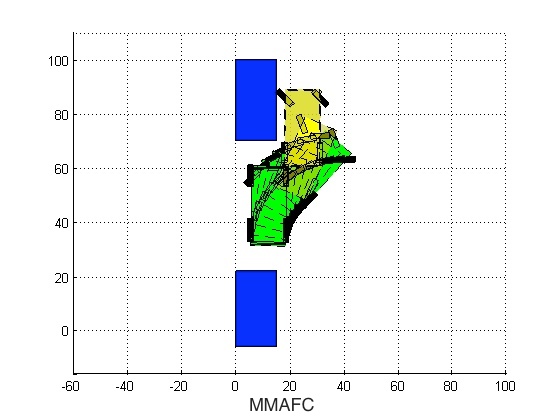}
\end{array}$
     \caption{Performance Comparison for Different Controllers}
     \label{Performance for Different Controllers}
   \end{figure}

\begin{figure}[thpb]     
     \centering
    \includegraphics[width=8cm,height=4cm]{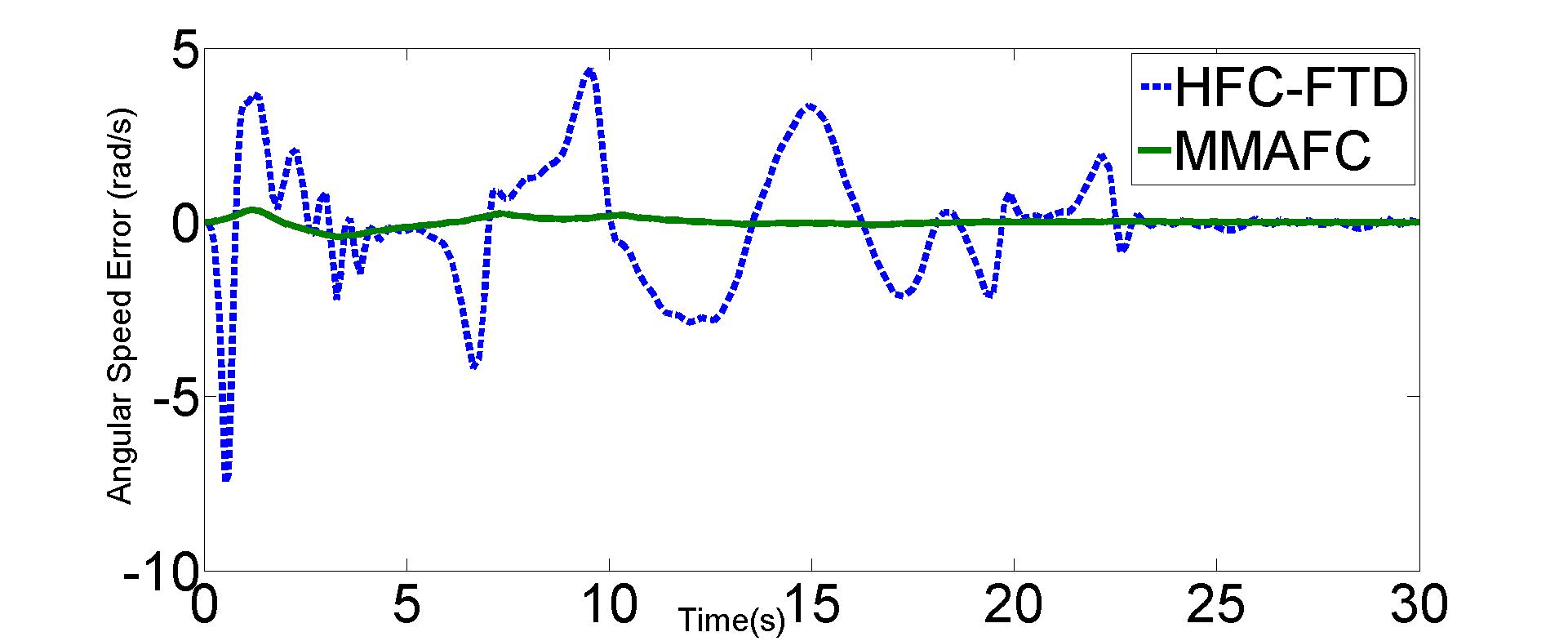}
    \caption{Tracking Error of Angular speed}
     \label{angular speed}
   \end{figure}

The simulation result is shown in Figure (\ref{Performance for Different Controllers}). The top two figures are for FBOS and HFC (from left to right), it is obvious that both of the controllers cannot fully complete the parking process due to a path distortion under a large ground friction. Comparatively, in the bottom two figures, it is shown that both cars controlled by HFC-FTD (left) and MM-AFLC (right) can complete the entire parking process though the procedure of MM-AFLC is much faster and smoother as indicated by those more scattered dots. Also the turning trajectory of the vehicle controlled by MM-AFLC is much smoother than that controlled by HFC-FTD due to a shorter transient period. The convergence of angular speed shown in Figure (\ref{angular speed}) also proves that MM-AFLC is more robust and gives superior performance than HFC-FTD under a large ground friction condition.

\section{Conclusion and Future Work}

In this paper, an Multiple Models Adaptive Fuzzy Logic Controller (MM-AFLC) is proposed for optimal autonomous control of an unmanned vehicle. As an extension of previous work in \cite{XYZ2012}, \cite{XYZ2013} and \cite{XYZ2014} on the same topic, The focus is on controller design to achieve better performance. The rule-base reasoning nature of a fuzzy logic controller ensures the system convergence under external disturbances. The process of on-line adaptation further improves the robustness while at the same time reduce efforts in controller tuning. Multiple models approach further enhances the system performance with faster convergence and less oscillation. By introducing neural network model identification, variations resulting from uncertainties in plant model can be minimized. As a result, the proposed controller can be used in different systems with minimal changes. A detailed stability analysis of the proposed system is also discussed in the paper.

It shall be noticed that there is an error between the actual fuzzy control signal and the optimal value . Although the error is bounded, it can negatively impact the overall system performance. Hence one important area of future improvement is to minimize the bounded error by adjusting the adaptive laws. Another potential research area is optimization for real industrial application. More practical challenges should be addressed such as implementation cost, operational efficiency and so on. 


\end{document}